\documentclass[twoside]{LCWS11}
\usepackage[latin1]{inputenc}
\usepackage[dvips]{graphicx,epsfig,color}
\usepackage{wrapfig,rotating}
\usepackage{amssymb,amsmath,array}
\usepackage{cite}
\usepackage{multirow}

\begin{document}
\title{Measurement of Chargino and Neutralino \\ Production at CLIC}
\author{Philipp Roloff$^1$, Timothy L. Barklow$^2$ and Astrid M\"{u}nnich$^1$
\vspace{.3cm}\\
1- CERN \\
1211 Geneva 23 - Switzerland \\
\vspace{.1cm}\\
2- SLAC \\
Stanford, CA 94309 - USA \\
}

\maketitle

\begin{abstract}
We present a study performed for the CLIC CDR on the measurement of chargino and neutralino production at $\sqrt{s} = 3$~TeV. Fully hadronic final states with four jets and missing transverse energy were considered. Results obtained using full detector simulation for the masses and for the production cross sections of the changino and the lightest and next-to-lightest neutralinos are discussed. 
\end{abstract}

\section{Introduction}

For the CLIC Conceptual Design Report (CDR)~\cite{ref:clic_cdr}, several physics processes were selected to benchmark the performance of general purpose detectors at a centre-of-mass energy of 3~TeV. The study presented in this document assumes a SUSY model defined by the mSUGRA parameters $m_{1/2} = 800$~GeV, $A_{0} = 0$, $m_{0} = 966$~GeV, $\tan\beta = 51$ and $\mu > 0$. In this model, the lightest chargino, $\tilde{\chi}^{\pm}_{1}$, has a mass of $643.2$~GeV, while the masses of the lightest and next-to-lightest neutralinos, $\tilde{\chi}^{0}_{1}$ and $\tilde{\chi}^{0}_{2}$, are given by $340.3$~GeV and $643.1$~GeV, respectively. The lightest neutral Higgs boson, $h^{0}$, has a mass of $118.5$~GeV.

The pair production of charginos and neutralinos was investigated:
\begin{equation}
e^{+}e^{-} \to \tilde{\chi}^{+}_{1}\tilde{\chi}^{-}_{1} \to W^{+}\tilde{\chi}^{0}_{1}W^{-}\tilde{\chi}^{0}_{1}~\textnormal{and}
\end{equation}
\begin{equation}
e^{+}e^{-} \to \tilde{\chi}^{0}_{2}\tilde{\chi}^{0}_{2} \to h^{0}(Z^{0})\tilde{\chi}^{0}_{1}h^{0}(Z^{0})\tilde{\chi}^{0}_{1},
\end{equation}
where $BR(\tilde{\chi}^{\pm}_{1} \to W^{\pm}\tilde{\chi}^{0}_{1}) = 100\%$, $BR(\tilde{\chi}^{0}_{2} \to h^{0}\tilde{\chi}^{0}_{1}) = 90.6\%$ and $BR(\tilde{\chi}^{0}_{2} \to Z^{0}\tilde{\chi}^{0}_{1}) = 9.4\%$. Hadronic decays of the $W^{\pm}$, $h^{0}$ and $Z^{0}$ bosons were considered and hence the investigated final state is given by four quarks and missing transverse energy. The reconstruction of chargino and neutralino pair production allows to benchmark the reconstruction of hadronically decaying gauge bosons in multi-hadron final states.

More details on the study discussed in the following are given in~\cite{ref:gaugino_analysis_clic_cdr}.

\section{Monte Carlo production}

The physics events used for the study presented here were generated using the WHIZARD 1.95~\cite{ref:whizard} program. Initial and final state radiation (ISR and FSR) were enabled during the event generation. The luminosity spectrum expected at CLIC was used during the event generation. The hadronisation of final state partons was simulated using PYTHIA~\cite{ref:pythia}. The generated events were accordingly passed through the detector simulation program SLIC which is based on the Geant4~\cite{ref:geant4} package. The CLIC\_SiD\_CDR~\cite{ref:clic_sid_cdr} detector geometry model was used.

Events were overlayed with pileup from $\gamma\gamma \to \textnormal{hadrons}$ interactions corresponding to 60 bunch crossings~\cite{ref:overlay_driver}. The reconstruction chain included an improved version of the PandoraPFA~\cite{ref:pandora_pfa} algorithm to reconstruct particle flow objects.
 
\begin{table}[!h]
\begin{center}
\begin{tabular}{|c|c|c|c|c|}\hline
Type &  Process & Cross section [fb] & Luminosity [ab$^{-1}$]& Referenced with\\ \hline
\multirow{2}{*}{Signal} &       $\tilde{\chi}_{1}^{+}\tilde{\chi}_{1}^{-} $         & 10.6 & 13.4 & Chargino\\
 &       $\tilde{\chi}_{2}^{0}\tilde{\chi}_{2}^{0} $         & 3.3 & 23.8 & Neutralino\\\hline
\multirow{7}{*}{Background} &   $\tilde{\chi}_{2}^{+}\tilde{\chi}_{2}^{-} $         & 10.5 & 1.8 & \multirow{4}{*}{SUSY}\\
 &       $\tilde{\chi}_{1}^{+}\tilde{\chi}_{2}^{-} $         & 0.8 & 8.9 & \\
 &       $\tilde{\chi}_{1}^{+}\tilde{\chi}_{1}^{-}\nu\overline{\nu}$    & 1.4 & 21.9 & \\
 &       $\tilde{\chi}_{2}^{0}\tilde{\chi}_{2}^{0}\nu\overline{\nu}$    & 1.2 & 13.3 & \\\cline{2-4}
 &       q$\overline{\text{q}}$q$\overline{\text{q}}\nu\overline{\nu}$                        & 95.4 & 4.5 & \multirow{3}{*}{SM}\\
 &       q$\overline{\text{q}}h^{0}\nu\overline{\nu}$                       & 3.1 & 6.2 &  \\
 &       $h^{0}h^{0}\nu\overline{\nu}$                      & 0.6 & 22.8 &\\\hline
\end{tabular}
\caption{Cross sections and integrated luminosities of the available Monte Carlo samples for chargino and neutralino pair production and for SUSY and Standard Model backgrounds. The charge conjugated modes are implied throughout this document.\label{tab:MCProdStats}}
\end{center}
\end{table}

An overview of all produced Monte Carlo (MC) samples is given in Tab.~\ref{tab:MCProdStats}. Dedicated samples for the considered signals corresponding to large luminosities are available. Additionally, backgrounds from SUSY and Standard Model (SM) processes were used in the study presented in this note.

\section{Event reconstruction}

The steps to reconstruct events with four jets from particle flow objects (PFOs) are described in this section. The presence of pileup from the process $\gamma\gamma \rightarrow$~hadrons increases the number of reconstructed PFOs in typical signal events by a factor 10 and the total visible momentum by a factor four. On the other hand, the background particles are emitted mostly in the forward direction. A large fraction of the background was rejected using combined timing and momentum cuts.

Only events containing at least four reconstructed PFOs with $p_{T} > 250$~MeV were considered further. Events with at least one identified electron or muon with $p_{T} > 20$~GeV were rejected.

Jets were reconstructed from PFOs using the $k_{t}$ algorithm~\cite{ref:kt_algorithm} as implemented in FastJet~\cite{ref:fastjet} in its exclusive mode with R = 0.7 and using the E recombination scheme. The clustering was stopped when four jets were found. To reject leptonic decays of $W^{\pm}$, $Z^{0}$ or Higgs bosons further, all jets were required to contain more than one PFO.

Bosons candidates were formed from jet pairs minimising:
\begin{equation}
(M_{jj,1} - M_{W^{\pm},h^{0}})^{2} + (M_{jj,2} - M_{W^{\pm},h^{0}})^{2},
\end{equation}
where $M_{jj,1}$ and $M_{jj,2}$ are the masses of the two reconstructed jet pairs and $M_{W^{\pm},h^{0}}$ was set to the world average of the $W^{\pm}$ boson mass to reconstruct $\tilde{\chi}_{1}^{\pm}$ and to the assumed Higgs boson mass to reconstruct $\tilde{\chi}_{2}^{0}$.

\begin{figure}[!h]
\begin{center}
\includegraphics[width=0.48\textwidth]{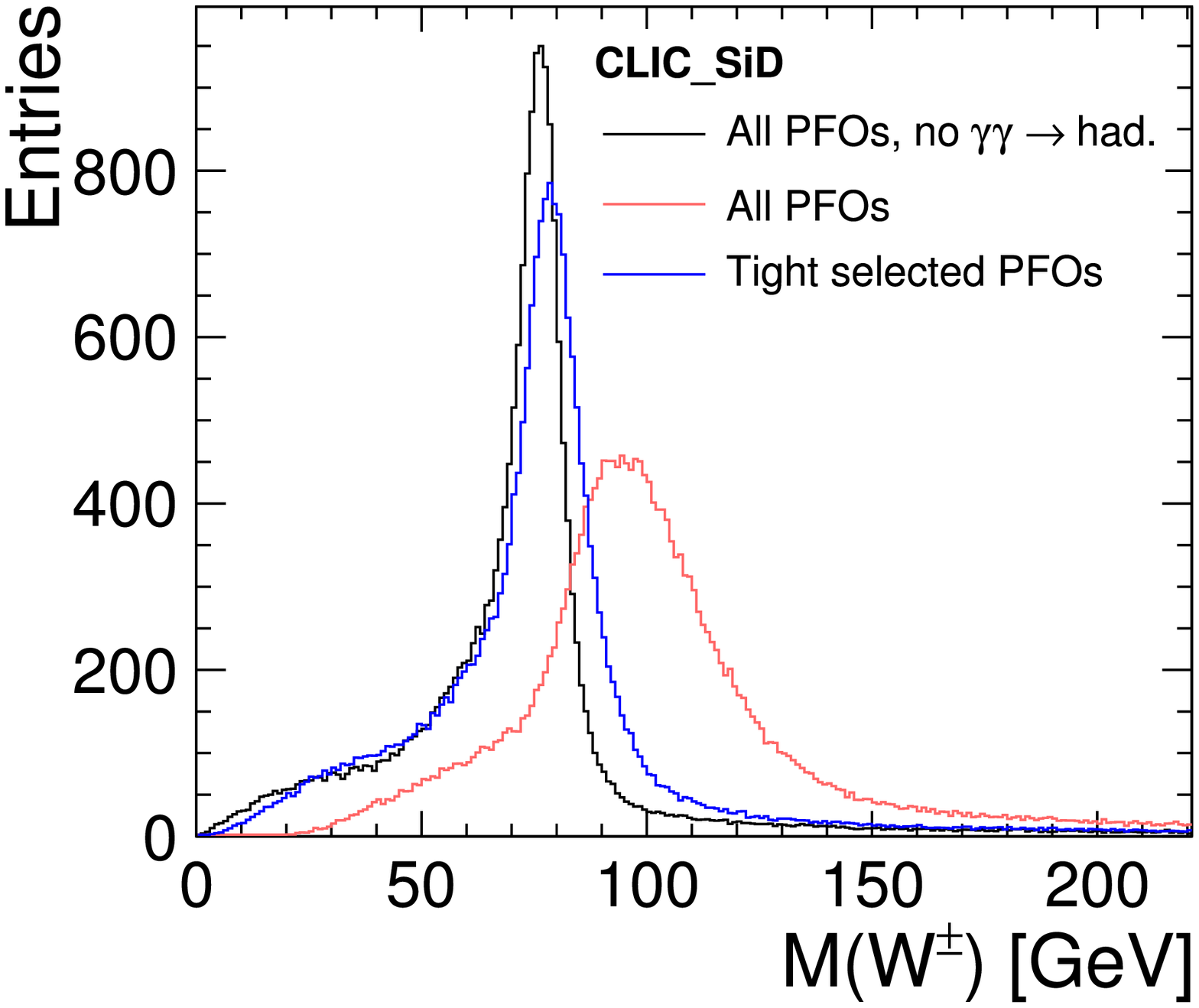}
\includegraphics[width=0.48\textwidth]{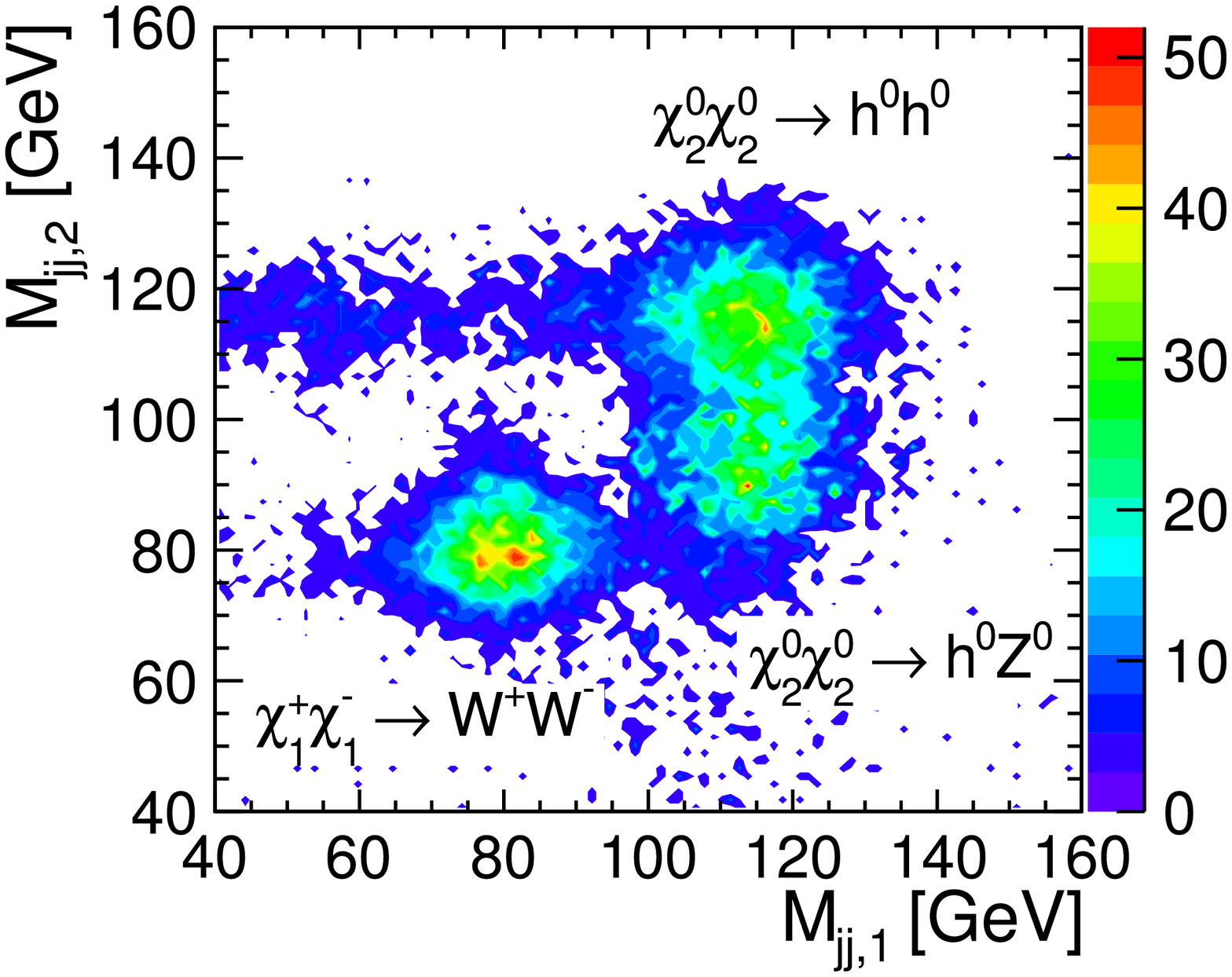}
\caption{Reconstructed mass of $W^{\pm}$ candidates without and with pileup from $\gamma\gamma \rightarrow $~hadrons (left). For events with pileup, the distribution for all PFOs is compared to that obtained using combined timing and momentum cuts (``Tight selected PFOs''). 2D mass plot of the signal final states for different decays of chargino and neutralino pairs (right). The peaks corresponding to the individual chargino and neutralino decays are indicated. The event samples were scaled to have similar number of events for each channel.\label{fig:W_reco_mass}}
\end{center}
\end{figure}

The reconstruction of $W^{\pm}$ bosons in $\tilde{\chi}_{1}^{+}\tilde{\chi}_{1}^{-}$ events is illustrated in Fig.~\ref{fig:W_reco_mass}. The distributions obtained with and without the overlay of $\gamma\gamma \rightarrow$~hadrons are compared. A good reconstruction of $W^{\pm}$ bosons was achieved if combined timing and momentum cuts were applied to select the PFOs used as input to the jet reconstruction. The right plot in Fig.~\ref{fig:W_reco_mass} shows the reconstructed masses for the two selected signal data sets: chargino pairs decaying into $W^{+}W^{-}$ and neutralino pairs decaying into either $h^{0}h^{0}$ or $h^{0}Z^{0}$. Since less than 1\% of the neutralino pairs decay to the $Z^{0}Z^{0}$ final state, this contribution is not shown in the figure. The horizontal band for $M_{jj,2} \approx M_{h^{0}}$ and $M_{jj,1} < M_{h^{0}}$ is caused by $\tilde{\chi}^{0}_{2}\tilde{\chi}^{0}_{2} \to h^{0}\tilde{\chi}^{0}_{1}h^{0}\tilde{\chi}^{0}_{1}$ events where one of the $h^{0}$ bosons is only partially reconstructed. No similar vertical band is visible due to the way the jets are ordered in the analysis.

\section{Event selection}

The selection of $\tilde{\chi}^{+}_{1}\tilde{\chi}^{-}_{1}$ and $\tilde{\chi}^{0}_{2}\tilde{\chi}^{0}_{2}$ pair production events was performed in two steps. First, a cut-based preselection was applied. The remaining backgrounds were suppressed further using boosted decision trees in a second step. These two steps are described in the following two subsections.

\subsection{Preselection cuts}
  
To restrict the training of the boosted decision trees to the region where the signal purities are high, the following preselection cuts were applied:
\begin{itemize}
\item $40 < M_\text{jj,1} < 160$~GeV~and~$40 < M_\text{jj,2} < 160$~GeV
\item $|\cos\theta^\text{miss}| < 0.95$, where $\theta^\text{miss}$ is the polar angle of the missing momentum
\item Angle between the $W^{\pm}$ or Higgs candidates larger than 1 radian
\item $|\cos\theta^\text{jj,1}| < 0.95$~and~$|\cos\theta^\text{jj,2}| < 0.95$, where $\theta^\text{jj,1}$ and $\theta^\text{jj,2}$ are the polar angles of the two jet pairs
\end{itemize}

\subsection{Event selection using boosted decision trees}

To distinguish between signal and background events further, boosted decision trees as implemented in TMVA~\cite{ref:tmva} were used. For training purposes, 20\% of the available events for each process were used. These events were not considered in the analysis to measure masses or cross sections.

The boosted decision trees were trained using 15 variables describing the event topology and describing kinematic quantities of the reconstructed $W^{\pm}$ or Higgs candidates. The efficiencies of the entire selection chain consisting of the preselection and of the BDT cut for reconstructed chargino and neutralino signal events were 25\% and 33\%, respectively. The signal purities in the selected samples were 57\% for the chargino and 55\% for the neutralino.

\begin{figure}[!h]
\begin{center}
\includegraphics[width=0.42\textwidth]{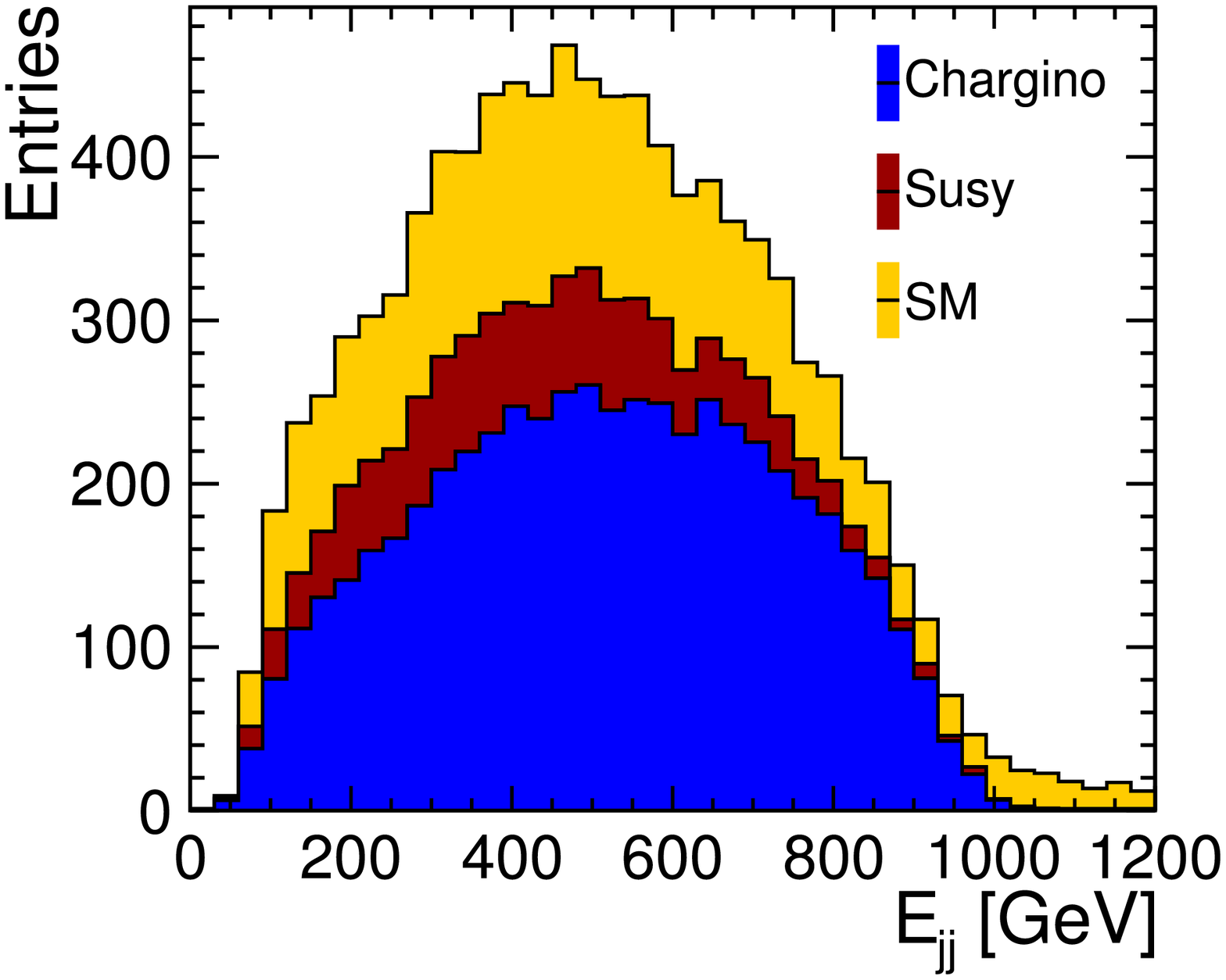}
\includegraphics[width=0.42\textwidth]{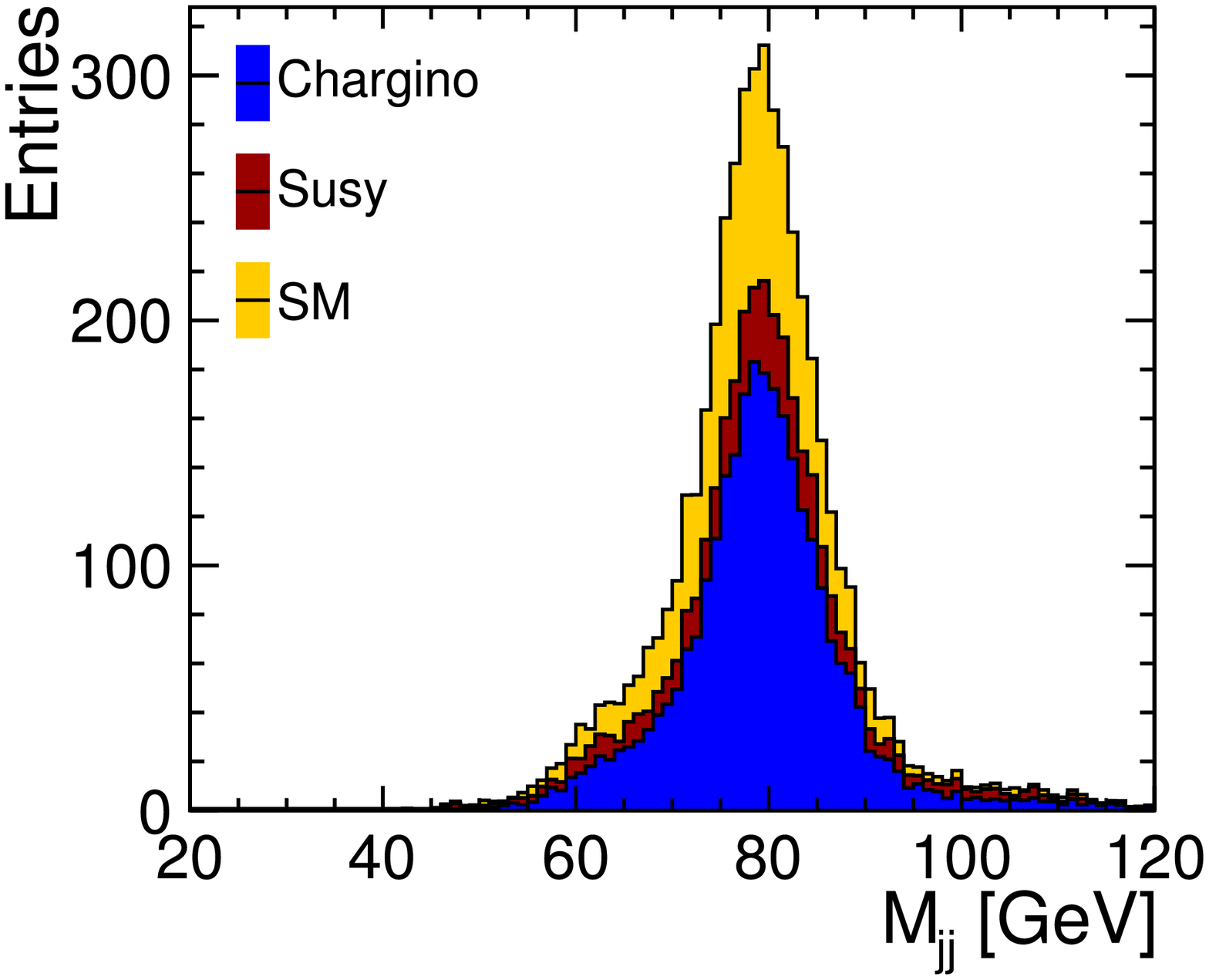}
\caption{Reconstructed $W^{\pm}$ energy (left) and mass (right) for the chargino signal and for the SM and SUSY backgrounds. The histograms are stacked on top of each other. All distributions are scaled to an integrated luminosity of 2~ab$^{-1}$.\label{fig:chargino_sel}}
\end{center}
\end{figure}

\begin{figure}[!ht]
\begin{center}
\includegraphics[width=0.42\textwidth]{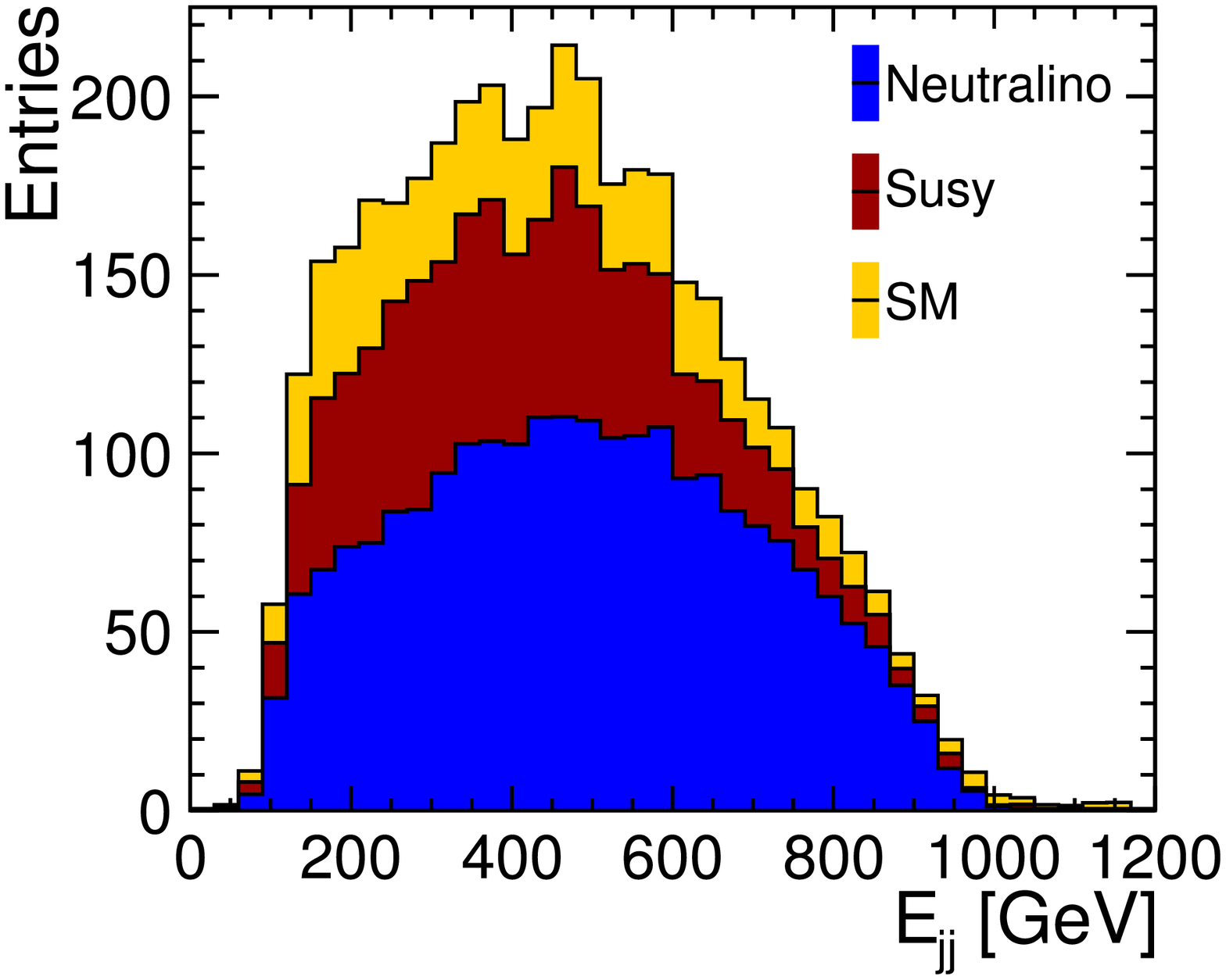}
\includegraphics[width=0.42\textwidth]{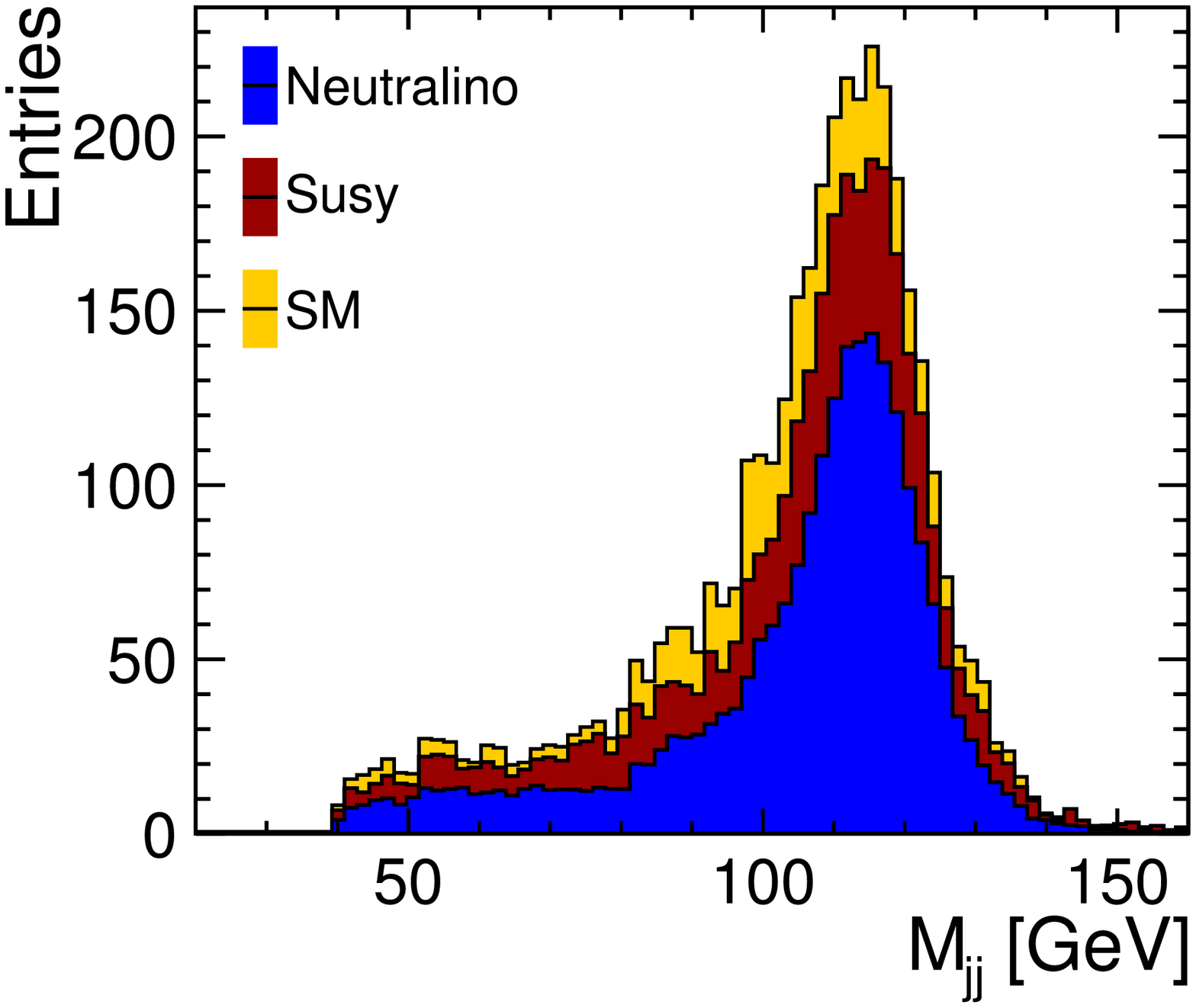}
\caption{Reconstructed Higgs energy (left) and mass (right) for the neutralino signal and for the SM and SUSY backgrounds. The histograms are stacked on top of each other. All distributions are scaled to an integrated luminosity of 2~ab$^{-1}$.\label{fig:neutralino_sel}}
\end{center}
\end{figure}

The energy and mass distributions of reconstructed Chargino and Neutralino candidates are shown in Figs.~\ref{fig:chargino_sel} and~\ref{fig:neutralino_sel}.

\section{Signal extraction}

Two complementary methods were used to extract masses and cross sections from the energy distributions of the reconstructed and selected $W^{\pm}$ and Higgs candidates (see Figs.~\ref{fig:chargino_sel} and~\ref{fig:neutralino_sel}). Both approaches are described in the following subsections.

\subsection{Template fit}

The pair production cross sections and masses of the $\tilde{\chi}_{1}^{\pm}$ and $\tilde{\chi}_{2}^{0}$ particles were determined using a template method where chargino and neutralino signal Monte Carlo samples for different mass hypotheses were generated. The $M(\tilde{\chi}_{1}^{0})$ mass was also measured since the energy distribution of $W^{\pm}$ bosons from $\tilde{\chi}_{1}^{\pm}$ decays is sensitive to this observable. Two-dimensional fits were performed simultaneously to the mass and production cross section for a given particle to account for the correlation between both quantities. The statistical uncertainties of the extracted masses and cross sections, determined using toy Monte Carlos, are shown in Table~\ref{tab:template_fit}. All measured values are in agreement with the input values used in the Monte Carlo generation.

\begin{table}
\centering
\begin{tabular}{c c c c}
\hline
 Parameter 1                & Uncertainty &          Parameter 2 & Uncertainty \\
\hline
 $M(\tilde{\chi}_{1}^{\pm})$ & $6.3$~GeV & $\sigma(\tilde{\chi}_{1}^{+}\tilde{\chi}_{1}^{-})$  & $2.2$\% \\
 $M(\tilde{\chi}_{1}^{0})$   & $3.0$~GeV & $\sigma(\tilde{\chi}_{1}^{+}\tilde{\chi}_{1}^{-})$  & $1.8$\% \\
 $M(\tilde{\chi}_{2}^{0})$   & $7.3$~GeV & $\sigma(\tilde{\chi}_{2}^{0}\tilde{\chi}_{2}^{0})$  & $2.9$\% \\
\hline
\end{tabular}
\caption{Uncertainties of the chargino and neutralino masses and pair production cross sections obtained from two parameter template fits. An integrated luminosity of $2$~ab$^{-1}$ is assumed.\label{tab:template_fit}}
\end{table}

\subsection{Least squares fitting}

Linear least squares fits of gaugino masses and cross sections were used to check the results of the template fitting technique and to fit for three or more  parameters simultaneously. Each $W^{\pm}/Z^{0}/h^{0}$ reconstructed energy histogram bin was expanded linearly about the nominal masses and cross sections. The slopes were obtained by convoluting a map of true-to-reconstructed bin contents with the true energy distributions at different gaugino masses. No fits were actually performed; instead the bin statistical errors were calculated and then propagated to the fit parameter errors using standard formulae for linear least squares fits.

The least squares results for two parameter fits are in reasonable agreement with the template fits described above. The two parameter fits assume that the other SUSY parameters have been measured with arbitrary accuracy. The $\tilde{\chi}_{1}^{0}$ mass will be measured with an accuracy of $\Delta M(\tilde{\chi}_{1}^{0}) = 3$~GeV at CLIC by combining the results from the slepton analyses~\cite{ref:slepton_analysis_clic_cdr}. A term constraining the $\tilde{\chi}_{1}^{0}$ mass to be within 3~GeV of the best estimate was added to the least squares fit. One can combine the data from the $\tilde{\chi}_{1}^{\pm}$ and $\tilde{\chi}_{2}^{0}$ analyses and perform a five parameter least squares fit of $M(\tilde{\chi}_{1}^{\pm}),M(\tilde{\chi}_{2}^{0}),M(\tilde{\chi}_{1}^{0}),\sigma(\tilde{\chi}_{1}^{+}\tilde{\chi}_{1}^{-}),\sigma(\tilde{\chi}_{2}^{0}\tilde{\chi}_{2}^{0})$. The results for this fit are shown in Tab.~\ref{tab:least_squares}.

\begin{table}
\centering
\begin{tabular}{c c}
\hline
 Parameter & Error \\
\hline
 $M(\tilde{\chi}_{1}^{\pm})$ & $7.3$~GeV \\
 $M(\tilde{\chi}_{1}^{0})$ & $2.9$~GeV \\
 $M(\tilde{\chi}_{2}^{0})$ & $9.8$~GeV \\
 $\sigma(\tilde{\chi}_{1}^{+}\tilde{\chi}_{1}^{-})$ & $2.4 \%$\\
 $\sigma(\tilde{\chi}_{2}^{0}\tilde{\chi}_{2}^{0})$ & $3.2 \%$ \\
\hline
\end{tabular}
\caption{Statistical mass errors and relative cross section errors for a five parameter least squares fit which includes the constraint that the $\tilde{\chi}_{1}^{0}$ mass be within 3~GeV of the value measured from slepton analyses.\label{tab:least_squares}}
\end{table}

\section{Summary}

The signals from $\tilde{\chi}_{1}^{\pm}$ and $\tilde{\chi}_{2}^{0}$ pair production were extracted from fully hadronic final states with four jets and missing transverse energy. The study was performed using full simulation and considering pileup from $\gamma\gamma \to \textnormal{hadrons}$. Two different signal extraction procedures are in reasonable agreement. The chargino and neutralino pair production cross sections were extracted with a statistical precision of $2 - 3\%$ while the masses of the $\tilde{\chi}_{1}^{\pm}$, $\tilde{\chi}_{1}^{0}$ and $\tilde{\chi}_{2}^{0}$ particles were determined with typical statistical accuracies of about $1 - 1.5\%$.

\begin{footnotesize}

\end{footnotesize}

\end{document}